\documentclass[final]{svjour3}
\usepackage{graphicx}
\usepackage{rotating}
\usepackage{amssymb}
\usepackage{mathptmx}
\usepackage[numbers]{natbib}
\usepackage{multirow}
\usepackage{caption}
\usepackage{courier}
\usepackage{subcaption}
\makeatletter
\journalname{Journal of Low Temperature Physics}

\bibpunct{[}{]}{,}{n}{}{,}

\begin{document}
\newcommand{\hdblarrow}{H\makebox[0.9ex][l]{$\downdownarrows$}-}
\title{The Thermal Conductance of Sapphire Ball Based Detector Clamps}

\author{H.D.~Pinckney \and G.~Yacteen \and A.~Serafin \and S.A.~Hertel \and For the Ricochet~Collaboration}

\institute{Department of Physics, University of Massachusetts Amherst,\\ Amherst, MA 01003, USA\\
\email{hpinckney@umass.edu}}

\maketitle

\begin{abstract}

In order to provide secure clamping with a low thermal conductance, some low temperature detectors are held with point contact sapphire ball clamps.  While this method is increasingly common, the thermal conductance across this interface has not been well studied.  We present a direct measurement of the thermal conductance of such clamps between 200 and 600~mK, with a clamping force of approximately 2.7~$\pm0.3$~N/clamp.  The thermal conductance of a single sapphire-on-copper clamp was found to be 660$^{+360}_{-210}$~$(T/K)^{3.1}$~[nW/K].  For a sapphire-on-silicon clamp the conductance was found to be 380$^{+190}_{-120}$~$(T/K)^{2.8}$~[nW/K].  The conductance measured is consistent with thermal boundary resistance.

\keywords{Thermal conductance, Sapphire ball clamping, detector holder}

\end{abstract}

\section{Introduction}

The challenge of rigidly holding low-temperature sensors while thermally isolating them from the environment typically employs insulating materials of small dimension.  Sapphire is commonly used as an insulating material~\cite{bintley,yoo,locatelli,suzuki}, and one common technique employs sapphire balls~\cite{ricochet}.  This method takes advantage of the low surface area of the contact between the sphere and the (typically planar) sensor surface to lower the conductance.  Additionally, the ball clamp geometry allows for a lower stress contact compared to a cone-based clamp, which helps to avoid inducing micro-fractures in the detector material.  While this clamping method is increasingly common, the exact thermal conductance has not been well documented in the literature.  In this article we present a study of the thermal conductance of such clamps.  While this study employs clamps replicating a Ricochet~collaboration holding style, the result is generally relevant to any sapphire ball clamp.

\section{Experimental Setup and Procedure}

\begin{figure}[htbp]
\begin{center}
\includegraphics[width=0.6\linewidth, keepaspectratio]{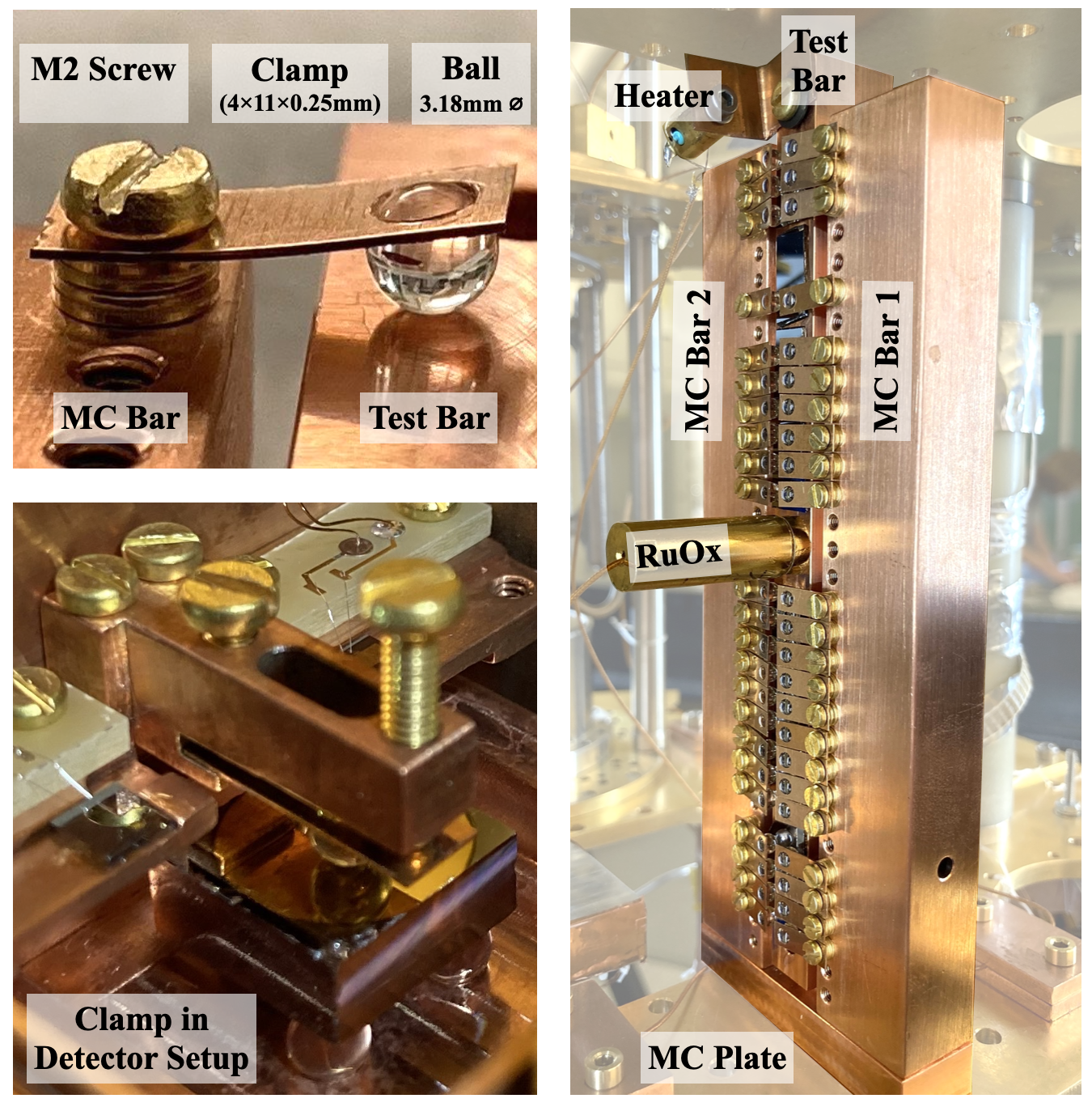}
\caption{Experimental setup.  \textit{Top Left:} Photograph of a single ball clamp during installation. \textit{Right:} Photograph of the entire assembly.  The central copper test bar is held by many clamps (half are on the back side, not visible here).  A single sapphire ball acts as a spacer at the bottom side, and it was observed to not make strong contact.  A heater and thermometer are fixed to the bar at top and middle respectively.  In this specific test, 92 sapphire ball based clamps are installed, and a polished Si surface has been installed between clamps and test bar. \textit{Bottom Left:} A single clamp as installed in a Ricochet R\&D detector housing.}
\label{fig:experiment}
\end{center}
\end{figure}

The experimental setup consists of an OFHC copper bar suspended by sapphire ball based clamps, see Fig.~\ref{fig:experiment}.  The bar was measured to have a surface roughness of 0.4 um R$_{a}$.  The clamps are based on 3.18 mm diameter sapphire balls with surface roughness less than 10~nm R$_{a}$~\cite{swissjewel}.  These balls are held in place by springs constructed of 4~mm~$\times$~11~mm~$\times$~0.25~mm hard-temper phosphor-bronze~\cite{goodfellow}.  The clamps are secured with brass screws and washers to an additional set of copper bars that act as an extension of the mixing chamber (MC) plate of a Cryoconcept Hexadry UQT-B 400 dilution refrigerator.  A thermal shield at 900~mK was installed for all measurements performed.  A Cryoconcept provided heater and RuOx thermometer were mounted on the bar and operated in a 4-wire measurements mode.  Signal connections were made with Manganin wire, shielded in a stainless steel braid with kapton insulation, and heat-sunk with copper clamps to the MC plate.  Data was recorded at 1 second intervals via the MMR3 bridge system and logged in the Cryoconcept software.  

In the following subsections we describe our determination of the clamping force, and then discuss our ``steady-state" and ``dynamic" data taking modes.

\subsection{Clamping Force}

The force of each clamp was estimated at room temperature with a spring scale, and cross-checked with Euler-Bernoulli beam theory.  Timoshenko beam theory was considered, however our system meets the assumptions required to apply the limiting case of Euler-Bernoulli theory.  We first describe the spring scale measurement, and then compare to the prediction from beam theory.

To perform the spring scale measurement, a clamp was installed on the bar.  A spring with no ball was installed directly next to this, and due to the absence of a ball was in the untensioned state.  We then used the spring scale to induce a deflection in the untensioned spring equal to that induced by the sapphire ball, as estimated by eye.  This was repeated five times across 2 different clamp setups.  Using this method, the force was determined to be 2.7~$\pm0.3$~N/clamp.

We now compare this result to prediction from beam theory.  Euler-Bernoulli beam theory describes the force of a bent beam as

\begin{equation}
    F_{clamp}=\frac{Ewt^3z}{4L^3}    \qquad  ,
    \label{eq:beam_theory}
\end{equation}

\noindent where $E$ is the Young's Modulus of the spring material, $w$ is the width of the spring, $t$ is the thickness of the spring, $z$ is the spring deflection distance, and $L$ is the length of the spring.  Using measurements of our springs and an assumed Young's Modulus of $100~\pm~10$~GPa~\cite{goodfellow} we get a force of approximately $2~\pm~1.2$~N/clamp.  This prediction is in agreement with the value from the spring scale measurement.

Last, the force is expected to change between room temperature and the operating temperature as a result of thermal contractions.  We bound this to be a 1\% effect based on measured thermal contraction results~\cite{nist_materials,pobell}.

\subsection{Conductance Measurement Methods}

Thermal conductance was measured in two complementary modes we'll refer to as ``steady-state" and ``dynamic".  

\subsubsection{Steady-State Mode}

In the steady-state measurement mode, we used the heater to supply power to the clamped bar and measured the temperature difference between the bar and the MC plate (\textit{i.e.}, across the clamps).  This assumes the clamps are the dominant thermal resistance between the bar and MC thermometers.  The thermal conductivity across the phosphor bronze spring and its brass fixture assembly was estimated to be orders of magnitude greater than the device under test~\cite{ekin}.

We assume a thermal conductance of the form~\cite{pobell}: 

\begin{equation}
	G = AT^{n} = \frac{dP}{dT}   \qquad   ,
\end{equation}

\noindent where $G$ is the thermal conductance, $T$ is the temperature of the material, $P$ is the total power through all clamps under test, and $A$, $n$ are constants.  Assuming this form we integrate to find a relationship between the temperature difference measured, the power applied, and the thermal conductance:

\begin{equation}
    P = P_{heater} + P_{external}= \frac{A}{n+1}\Big(T_{bar}^{n+1}-T_{MC}^{n+1}\Big)
    \label{eq:master_fit}
\end{equation}

\noindent where $T_{bar}$ is the temperature of the bar and $T_{MC}$ is the temperature of the MC.  We have separated the total power into the power applied with the heater ($P_{heater}$) and the incidental power from other external heat loads ($P_{external}$).  This external heat load is likely dominated by the wiring to the bar's heater and thermometry, which were estimated to have have at most a 10~nW heat load on the bar when the bar is at 200~mK and the MC is at 9~mK~\cite{pobell}.  The expected radiative heat load on the copper bar test mass is at most 0.3~nW.  This heat load is due to black body radiation from the thermal shield installed on our refrigerator's 0.9~K stage.  For the purpose of our fits, $P_{external}$ is treated as a constant, and the systematic uncertainty associated with this is discussed in section~\ref{analysis and results}.

\begin{figure}[htbp]
\begin{center}
\includegraphics[width=0.6\linewidth, keepaspectratio]{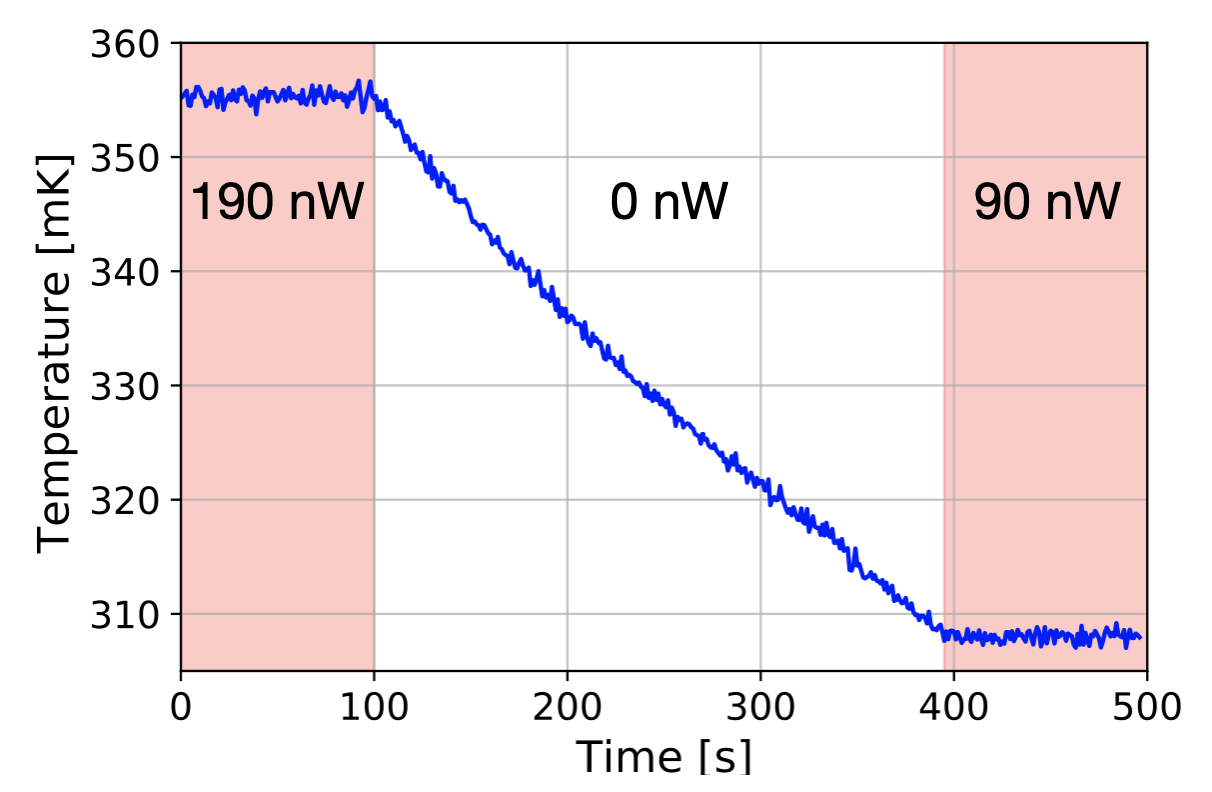}
\caption{Example of data collected in dynamic mode.  The dominant uncertainty in temperature is a 10~\% calibration offset, not illustrated here.  Power is on in shaded regions.  The power applied in each region is annotated. (Color figure online)}
\label{fig:dynamic_data}
\end{center}
\end{figure}

The data-taking procedure was to supply 10 different heater powers at a constant $T_{MC}$, measure $T_{bar}$ at each heater power, and fit for $A$, $n$, and $P_{external}$.  The heater power was varied between 15~nW and 1~$\mu$W, corresponding to bar temperatures between 200~mK and 600~mK.  We performed this measurement for a setup with both 48 and 92 clamps, and with $T_{MC}$ fixed at both 9~mK and 150~mK.  We also tested for a dependence on substrate material by inserting pieces of polished silicon wafer between the copper bar and sapphire balls.  The Si was thermally anchored to the Cu bar using GE 7031 Varnish, and 92 clamps were in place for this Si interface measurement.  The silicon-varnish-copper conductance was estimated to have a 10~mW/K conductance at 200~mK~\cite{pobell,si_conductivity}, many orders of magnitude greater than the clamps under test.  The noise in the thermometer was approximately 1~mK for each data point.

\subsubsection{Dynamic Mode}

The dynamic measurement method observes the time-dependent temperature response after a sudden change in applied power.  The temporal response depends on a combination of thermal conductance and heat capacity.  The temporal response can be used as a measure of thermal conductance in this case because the heat capacity of the bar can be well-estimated.

The measurement begins with the bar at near-constant temperature under constant applied power (as in the steady-state measurement).  The applied heater power is then abruptly turned off, and the bar's temperature is logged as a function of time for  a few minutes.  An example data trace is shown in Fig.~\ref{fig:dynamic_data}.  Such data can be described by the following differential equation:

\begin{equation}
    C\frac{dT}{dt} = BT\frac{dT}{dt} = \frac{A}{4}\Big(T_{bar}^{4}-T_{MC}^{4}\Big) - P_{external} \qquad   ,
\end{equation}

\noindent where $C=BT$ is the heat capacity of the copper test mass at low temperature~\cite{cu_heat_capacity}, $B$ is a constant, and we have set $n=3$ from Eq.~\ref{eq:master_fit}.

\section{Analysis and Results}
\label{analysis and results}

Steady-state fitting was performed using the python \texttt{iminuit} package.  Dynamic fitting was done using the \texttt{scipy.optimize.curve\_fit} package.  

\begin{figure}[htbp]
\begin{center}
\includegraphics[width=0.7\linewidth, keepaspectratio]{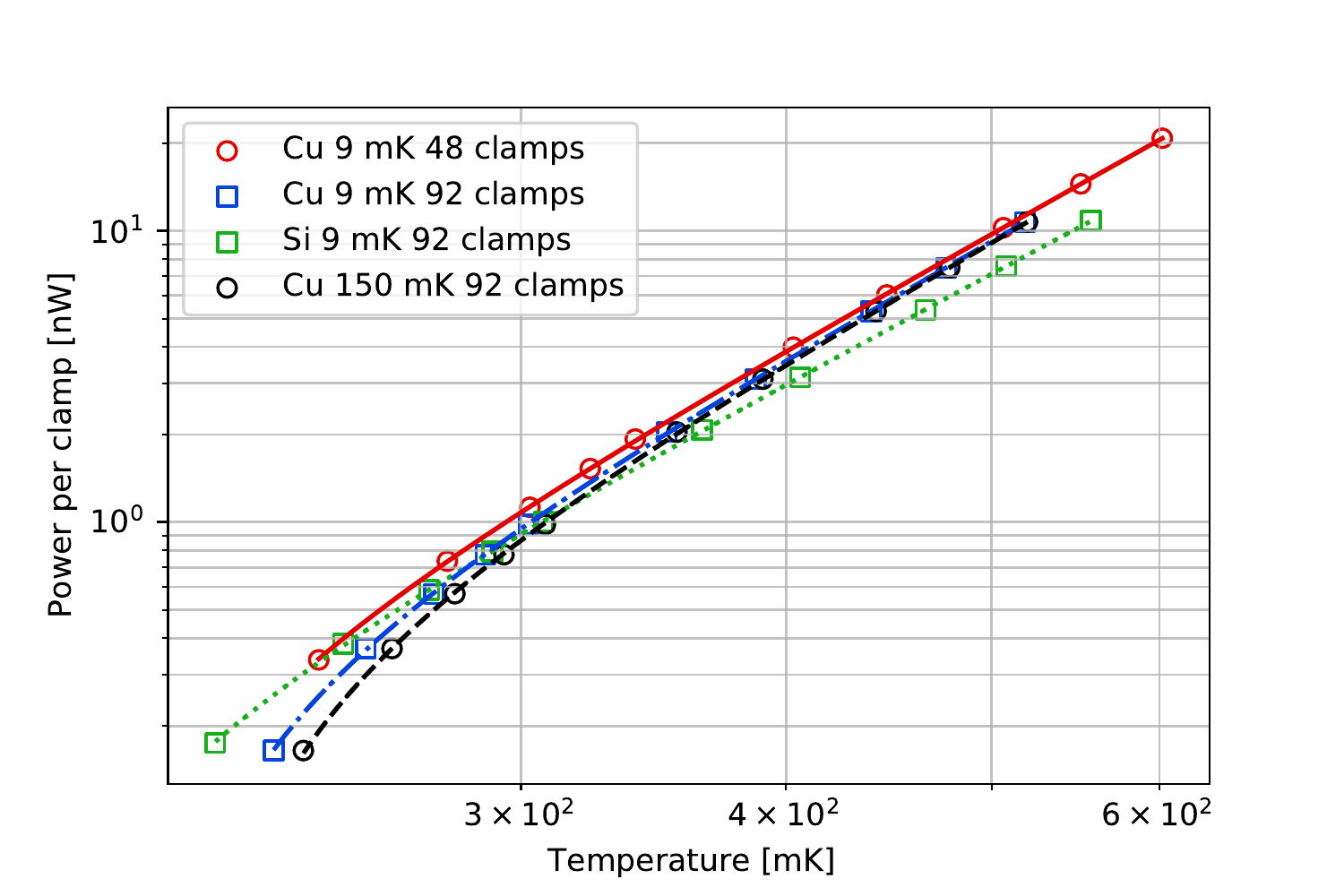}
\caption{Example fits to the steady-state data for datasets 0, 2, 3, and 4 in Fig.~\ref{fig:results}). Data is shown as markers, and the lines are the fit to the data.  The fit function is given by equation~\ref{eq:master_fit}.  (Color figure online.)}
\label{fig:steady_state_fit}
\end{center}
\end{figure}

Fig.~\ref{fig:steady_state_fit} shows the fit results for datasets with a variety of conditions, including varying the clamp number, substrate material, and MC temperature.  The fits qualitatively agree with the data.  Systematic uncertainties due to the temperature offset is not displayed.  Fixing the exponent to n$=3$, as would be motivated by bulk phonon transport or acoustic mismatch at the boundary, does not dramatically change the fit results for sapphire-on-copper data, though it does qualitatively worsen the fit at lower temperatures for the sapphire-on-silicon data  Allowing $P_{external}$ to have a temperature dependence has a few-percent effect on the amplitude and exponent.  For the amplitude, this uncertainty is sub-dominant to the temperature systematic.  For the exponent this uncertainty is at the level of the first decimal place.  Additionally, the effect of increasing $T_{MC}$ is apparent below approximately 300~mK, where $T_{bar}$ approaches the higher set point of $T_{MC}$.  

\begin{figure}[htbp]
\begin{center}
\includegraphics[width=0.8\linewidth, keepaspectratio]{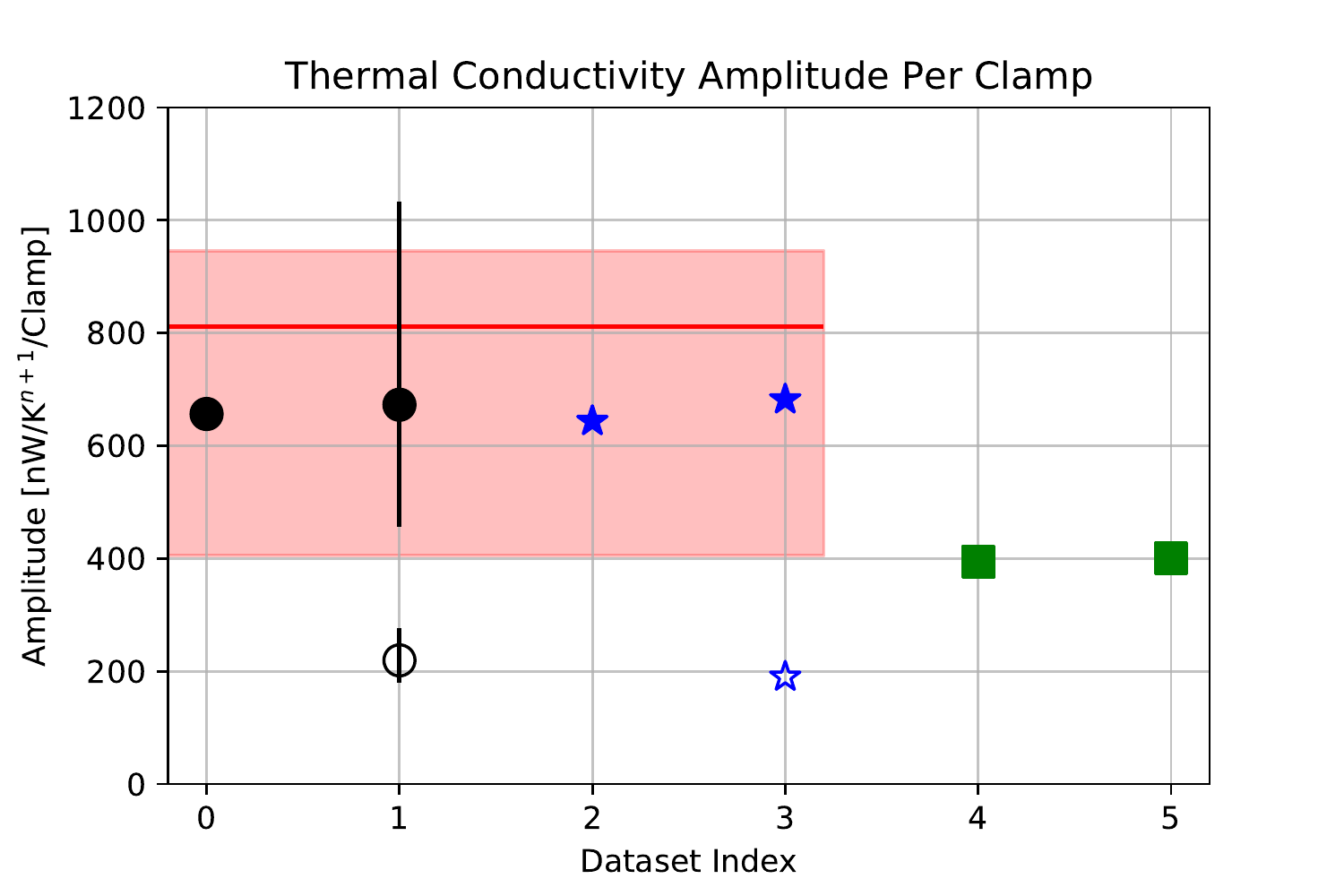}
\caption{Thermal conductance amplitude per clamp for the various data sets and selected uncertainties.  Black circles represent data taken with 48 clamps directly on copper.  Blue stars represent data taken with 92 clamps directly on copper.  Green squares represent data taken with 92 clamps on silicon.  Filled markers are the result of steady-state fits, hollow markers are the result of dynamic fits.  The red band indicates the acoustic mismatch model of combined thermal boundary resistance between copper-sapphire, and sapphire-phosphor-bronze spring, with uncertainty dominated by our area estimate.  Datasets 0, 2, and 4 were taken with the MC at 9~mK, and datasets 1, 3, and 5 were taken with the MC at 150~mK.  The uncertainties plotted are the systematic uncertainty given by scaling the thermometry by 10~\% in either direction.  Scaling the temperature down increased the conductance amplitude for steady-state and dynamic fits, and scaling the temperature up decreased the conductance amplitude.  Uncertainties on all points are nearly identical to dataset 1, and so for clarity only those for dataset 1 are plotted. (Color figure online.)}
\label{fig:results}
\end{center}
\end{figure}

Fig.~\ref{fig:results} compares the results of all measurements using both the steady-state and dynamic methods, and Table~\ref{tab:results} displays the average fit values for each method.  The per-clamp conductance amplitude agrees well between the 48 and 92 clamp tests.  Additionally, the change from a copper to a silicon interface surface decreased the thermal conductance by nearly a factor of 2.  This could be due to material hardness; we expect a smaller area of contact between the ball and the silicon.  Additionally, we notice that the steady-state and dynamic data sets are consistent to within a factor of approximately 3.  One improvement that could be made is a refined, more complex thermal model, which would allow for an additional heat capacity and thermal conductivity between the bar and bath.  In addition to improving the thermal model, a temperature dependent treatment of $P_{external}$ in the dynamic data fits could result in improved agreement between the dynamic and steady-state methods.

Last, we can use the acoustic mismatch theory presented in~\cite{schwartzpohl} to make a prediction of the thermal conductance due to boundary resistance and compare with these measured values.  We estimate from a scratch test that the contact area between the sapphire and copper was $3^{+0.5}_{-1.5}~\times~10^{-5}$~cm$^2$, and taking the boundary resistance between sapphire and copper to be 18.5~K$^4$/W$^{1}$/cm$^{2}$ from~\cite{schwartzpohl}, we can estimate that the conductance per clamp due to boundary resistance at the sapphire-copper interface should be 1600~$^{+300}_{-800}$~nW/K$^{4}$.  As phosphor bronze and copper have a nearly identical modulus of elasticity, we expect a similar area at the ring contact between the sapphire ball and the phosphor bronze spring.  As phosphor bronze and copper have similar accoustic properties, we then assume the clamp is well modeled by two Kapitza resistances in series, each of magnitude similar to the sapphire-copper interface resistance.  Therefore, the total thermal conductivity of the clamp is estimated to be 800~$^{+150}_{-400}$~nW/K$^{4}$.  This agrees with our measurement, and we attribute the thermal resistance to be dominated by boundary resistance.  Comparison between theory and experiment was not made for sapphire on silicon because of a large uncertainty in the area of contact.

\begin{table}[h!]
\begin{center}
\begin{tabular}{|c|l|l|l|}
\hline
\multirow{6}{*}{Steady-state} & \multirow{3}{*}{Cu} & A               & 660~nW/K$^{n+1}$/clamp \\ \cline{3-4} 
                              &                     & n               & 3.1 \\ \cline{3-4} 
                              &                     & P$_{external}$  & 16~nW \\ \cline{2-4} 
                              & \multirow{3}{*}{Si} & A               & 380~nW/K$^{n+1}$/clamp \\ \cline{3-4} 
                              &                     & n               & 2.8 \\ \cline{3-4} 
                              &                     & P$_{external}$  & 13~nW \\ \hline
\multirow{2}{*}{Dynamic}      & \multirow{2}{*}{Cu} & A               & 200~nW/K$^{4}$/clamp \\ \cline{3-4}
                              &                     & P$_{external}$  & 15~nW \\ \hline
\end{tabular}
\caption{Mean results of the fits.  Steady-state results are averaged over the 9~mK and 150~mK base temperatures.  Uncertainties in the exponents are in the first decimal place, and are dominated by the systematic uncertainty of the temperature-independent treatment of $P_{external}$.}
\label{tab:results}
\end{center}
\end{table}

\section{Conclusions}

We have measured the thermal conductance of sapphire ball clamps and have shown that the interface appears to be dominated by a thermal boundary resistance.  The thermal conductance was found to be 660$^{+360}_{-210}$~$(T/K)^{3.1}$~[nW/K] for sapphire on copper, and 380$^{+190}_{-120}$~$(T/K)^{2.8}$~[nW/K] for sapphire on silicon.  Additionally, in conjunction with simulations in~\cite{thermalchip}, this measurement shows that a detector can be held with 9 clamps of this style and not suffer significant resolution degradation.

\section{Data Availability}

The datasets analysed during the current study are available from the corresponding author on request.

\begin{acknowledgements}
This work was partially supported by DOE QuantISED award DE-SC0020181.  Template v.1 by KLL - June 18, 2015.
\end{acknowledgements}

\pagebreak


\begin{thebibliography}{99}

\bibitem{bintley}
D. Bintley, A.L. Woodcraft, F.C. Gannaway, {\it Cryogenics} \textbf{47}, p. 333-342 (2007), DOI: 10.1016/j.cryogenics.2007.04.004

\bibitem{yoo}
K-H. Yoo, A.C. Anderson, {\it Cryogenics} \textbf{23}, p. 531-532 (1983), DOI: 10.1016/0011-2275(83)90187-X

\bibitem{locatelli}
M. Locatelli, D. Arnaud, M. Routin, {\it Cryogenics} \textbf{16}, p. 374-375 (1976), DOI: 10.1016/0011-2275(76)90220-4

\bibitem{suzuki}
T. Suzuki et. al, {\it Proceedings of the 28th International Cosmic Ray Conference} p. 3131-3134 (2003), ADS: 2003ICRC....5.3131S

\bibitem{ricochet}
C.~Augier et al [Ricochet], {\it J. Low Temp. Phys.} This Special Issue (2021)

\bibitem{swissjewel}
Swiss Jewel part number B3.18S, https://www.swissjewel.com/

\bibitem{goodfellow}
Goodfellow part number CU050260, http://www.goodfellow.com/

\bibitem{nist_materials}
D.E. Apostolescu, P.S. Gaal, and A.S. Chapman, {\it A proposed high temperature thermal expansion reference material}, \textbf{Standard Reference Materials}, p.637-646, Accessed through NIST Materials Database 

\bibitem{pobell}
F. Pobell, {\it Matter and Methods at Low Temperatures}, Springer, Berlin, Heidelberg (2007), DOI:10.1007/978-3-540-46360-3

\bibitem{ekin}
J. Ekin, {\it Experimental Techniques for Low Temperature Measurements}, Oxford University Press, New York, (2006), DOI:10.1093/acprof:oso/9780198570547.001.0001

\bibitem{si_conductivity}
C.J. Glassbrenner, G.A. Slack, {\it Phys. Rev.}, \textbf{134}, A1058, (1964) DOI:10.1103/PhysRev.134.A1058

\bibitem{cu_heat_capacity}
R.J. Corruccini, J.J. Gniewek, {\it NBS Mono}, \textbf{21}, (1960), DOI:10.6028/NBS.MONO.21

\bibitem{schwartzpohl}
E.T. Swartz and R.O. Pohl, {\it Rev. Mod. Phys.} \textbf{61}, 605, (1989), DOI:10.1103/RevModPhys.61.605

\bibitem{thermalchip}
R.~Chen et al [Ricochet], {\it J. Low Temp. Phys.} This Special Issue (2021)

\end{thebibliography}
\end{document}